\documentclass{article}%
\usepackage{amsmath}
\usepackage{amsfonts}
\usepackage{amssymb}
\usepackage{graphicx}%
\providecommand{\U}[1]{\protect\rule{.1in}{.1in}}

\begin{document}

\title{Relay QKD Networks \& Bit Transport}
\author{\textit{Simon JD Phoenix}$^{1}$\textit{\ }\& \textit{Stephen M Barnett}%
$^{2}\bigskip$\\$^{1}$Khalifa University, PO Box 127788, Abu Dhabi, UAE\\$^{2}$School of Physics and Astronomy, University of Glasgow,\\Glasgow G12 8QQ, UK}
\maketitle

\begin{abstract}
We show how it is possible to operate end-to-end relays on a QKD network by
treating each relay as a \textit{trusted} eavesdropper operating an
intercept/resend strategy. It is shown that, by introducing the concept of
`bit transport', the key rate compared to that of single-link channels is
unaffected. The technique of bit transport extends the capability of QKD
networks. We also discuss techniques for reducing the level of trust required
in the relays. In particular we demonstrate that it is possible to create a
simple quantum key exchange scheme using secret sharing such that by the
addition of a single extra relay on a multi-relay channel requires the
eavesdropper to compromise all the relays on the channel. By coupling this
with multi-path capability and asynchronous quantum key establishment we show
that, in effect, an eavesdropper has to compromise all relays on an entire
network and collect data on the entire network from its inception.

\end{abstract}

\section{Introduction}

Quantum Key Distribution (QKD) is an elegant and ingenious application of
quantum mechanics to the problem of key establishment in classical
cryptosystems (see [1,2] for the 2 seminal works that established the two
basic methods of QKD; for some excellent reviews of QKD see [3]). Its
principal limitation, in practical terms, is the distance over which it is
possible to exchange the keys securely. Whilst the security of the technique
is robust to loss, detector imperfections mean that even modest amounts of
loss, such as that seen in optical fibres, can limit the distance of the
technique over a single channel. Any widespread implementation of QKD must, of
course, operate on a network, and this introduces further complications.
Active processing of any quantum signal will, in general, destroy the
integrity of the quantum states and render quantum key exchange impossible.
QKD works beautifully on passive optical networks [4,5], but passive network
switching elements can also be seen as effectively introducing further loss,
thus adding to the difficulty of operating a QKD network over the distances
that might be required in any realistic network implementation.

The current design of QKD network trials (see, for example, [6]) adopt an
approach based on relay nodes operated in a link-by-link fashion in order to
extend the distance (we use the terminology `LL relays' to describe such
link-by-link operation). Whilst this is an eminently practical solution to the
problem of extending the distance of any QKD scheme it is preferable, from a
security perspective, to have the capability of exchanging end-to-end keys
between any two network users. Bechmann-Pasquinucci and Pasquinucci [7] show
how, by considering a relay to be a \textit{trusted} eavesdropper operating a
standard intercept/resend strategy, it is possible to construct such an
end-to-end relay (we shall use the term `IR relays' to describe these). It is
argued in [7] that such relays cannot be used, however, to extend the distance
for a QKD channel. By a suitable adaptation of the operation of these IR
relays we show here how, contrary to this conclusion, it is indeed possible to
use such relays to extend the distance.

Both LL and IR relays must be \textit{trusted} network elements and this, to
some extent, reduces the attractiveness of the technology. After all, perhaps
the most compelling feature of single link QKD is to allow key exchange in a
provably-secure fashion. We discuss techniques for reducing the level of trust
required in such relay solutions. In particular, we show that by adding in
redundancy at the relay level, we can construct channels in which an
eavesdropper has to compromise \textit{all} of the relays. The implication of
this is that our network users, Alice and Bob, only need to be able to trust
\textit{at least one} relay on such channels\footnote{Some preliminary results
were presented at the \textit{IEEE GCC Conference}, Dubai, UAE $\left(
2011\right)  $}.

A naive application of IR relays would suggest that each relay reduces the
effective key rate by a factor of 2 from the key rate that can be achieved by
single link QKD. This exponential reduction in key rate as we add more relays
is catastrophic for our original goal of extending the distance. By
introducing the notion of bit transport we show that, in fact, the single link
QKD rate can be achieved over channels with any number of relays. Bit
transport is a powerful classical post-processing technique that can be used
to extend the flexibility and capability of QKD networks [8] and we discuss
some applications here, including the ability of Alice and Bob to perform
eavesdropper detection on a duplex QKD channel without the necessity of public
bit comparison.

\section{Intercept-Resend Relays}

An IR relay as envisioned in [7] acts precisely as an eavesdropper performing
an intercept/resend strategy on each transmitted quantum state. The relay
chooses at random one of the coding bases used by Alice for each timeslot and
simply retransmits the state corresponding to the measurement result. The
difference between an eavesdropper and the relay is that now we consider the
relay to be a cooperative entity on the network. We can also view the action
of the relay as another network user, such as Bob, whose function is to
re-transmit the measured result to the next network user. One of the nice
features of using relays in IR mode is that we do not have to consider a
single dedicated path between Alice and Bob, but can consider the
establishment of the shared key on multiple distinct paths between them on the
network. This feature alone allows us to put less trust in any individual path
and the relays on that path. When this is combined with the technique of
secret-sharing we can reduce the level of trust required in any individual
relay still further [9,10]. We shall discuss the issue of trust as it applies
to the relay nodes of a quantum network later.

We shall assume the coding bases used are represented by the operators
$\hat{X}$ and $\hat{Y}$ which have eigenstates $\left\vert \pm\right\rangle
_{X}$ and $\left\vert \pm\right\rangle _{Y}$, respectively, and we adopt the
coding interpretation $\left\vert +\right\rangle \equiv1$ and $\left\vert
-\right\rangle \equiv0$. Thus, the quantum key is to be established using the
BB84 protocol [1]. We shall label the relays on a channel by $R_{j}$ and the
two users who wish to establish a key, Alice and Bob, by $A$ and $B$,
respectively. The distance between $A$ and $B$ is such that a quantum key
cannot be successfully established by single link QKD. A channel between Alice
and Bob with a single relay is of the form%
\[
A\longrightarrow\fbox{$R$}\longrightarrow B
\]
The possible transmissions on the channel, when Alice chooses to use the
coding basis represented by $\hat{X}$, can be split into the 4 cases shown in
the table below%

\[%
\begin{tabular}
[c]{ccc}%
\underline{Alice} & \underline{\textit{R}} & \underline{Bob}\\
$X$ & $X$ & $X$\\
$X$ & $Y$ & $X$\\
$X$ & $X$ & $Y$\\
$X$ & $Y$ & $Y$%
\end{tabular}
\ \
\]
with a similar table for the situations in which Alice makes the choice of
coding basis represented by $\hat{Y}$. It is clear that a quantum key can be
established on the first of these entries in the table in which we have all 3
parties using the same coding basis. At first sight this suggests that an IR
relay can be used to extend the distance for key establishment, but at the
expense of sacrificing 3/4 of the transmissions, instead of the loss of 1/2 of
the transmissions on filtering that we would obtain for single link QKD.
However, if the channel between Alice and the relay is already operating at
the limit for single link QKD then simple re-transmission of the measured
results by the relay will lead to a quantum signal between the relay and Bob
that is insufficient to overcome the signal-to-noise limitations at Bob's
detector. Bechmann-Pasquinucci and Pasquinucci [7] prove that a relay channel
operated in this fashion cannot be used to extend the distance for QKD.

The argument presented in [7] is applicable to the situtation when we think of
the overall channel $A\rightarrow R\rightarrow B$ as a single channel. The
problem is that the final quantum signal that gets to Bob is too weak to be
successfully distinguished from the detector noise. However, we do not need to
operate this as a single channel, but can effectively split the communication
into 2 separate channels. In order to overcome the signal-to-noise problem at
Bob's detector we can do one of two things:

\begin{enumerate}
\item The relay delays retransmission until sufficient data has been gathered
so that the signal sent on to Bob is of sufficient strength to overcome the
SNR limitations at Bob's detector

\item The relay pads the retransmission with a separate QKD communication
between the relay and Bob. The relay can keep track of which timeslots are
from Alice and which are padding qubits. The padding qubits can be used to
establish a separate quantum key between the relay and Bob, if desired.
\end{enumerate}

Using either of the techniques (1) and (2) we see that we can use IR relays to
successfully extend the distance for quantum key establishment. The apparent
problem with this is that we have halved the key rate between Alice and Bob.
It is clear that the addition of another relay further exacerbates the
situation and each additional relay will introduce a further loss factor of 2
into the final key rate. However, this reasoning is based on the
\textit{assumption} that only the channels where \textit{all} entities have
used the same coding basis can be used to establish an end-to-end key between
Alice and Bob. As we shall see in the next section, the notion of bit
transport reveals that this is an unduly restrictive assumption and that the
introduction of relays does not affect the final key rate.

\section{Bit Transport}

In order to illustrate the technique of bit transport we shall consider a
channel that requires two relays, $R_{1}$ and $R_{2}$, to span the distance
between $A$ and $B$. Each relay is operated in IR mode as described above. The
transmission sequence is schematically described by%

\[
A\longrightarrow\fbox{$R_1$}\longrightarrow\fbox{$R_2$}\longrightarrow B
\]
In the timeslots where Alice chooses the coding basis $X$ we have the
following possible situations\footnote{For the sake of brevity we will simply
describe the choice of coding basis by $X$ or $Y$ rather than using the more
cumbersome form `the coding basis represented by $\hat{X}$', for example.}:%

\[%
\begin{tabular}
[c]{c|cccc}
& Alice & $R_{1}$ & $R_{2}$ & Bob\\\hline
&  &  &  & \\
1 & $X$ & $X$ & $X$ & $X$\\
2 & $X$ & $X$ & $X$ & $Y$\\
3 & $X$ & $X$ & $Y$ & $X$\\
4 & $X$ & $X$ & $Y$ & $Y$\\
5 & $X$ & $Y$ & $X$ & $X$\\
6 & $X$ & $Y$ & $X$ & $Y$\\
7 & $X$ & $Y$ & $Y$ & $X$\\
8 & $X$ & $Y$ & $Y$ & $Y$%
\end{tabular}
\]
We can think of the entire transmission (in which Alice chooses the coding
basis $X$) as being made up of 8 distinct channels, in which each channel
occurs at random in the sequence of timeslots. There is, as discussed
previously in the case of the single-relay channel, an equivalent table
describing the situations where Alice chooses the coding basis $Y$. Channel 3
in the above table, for example, represents those instances where relay 1
chooses the coding basis $X$, relay 2 the coding basis $Y$, and Bob chooses
the coding basis $X$.

In full, therefore, we have a complete transmission consisting of $N$
timeslots. On average, in $N/2$ of these Alice will choose the coding basis
$X$ and the coding basis $Y$ in the remaining timeslots. Thus the entire
transmission can be partitioned into 16 channels, distributed at random in the
timeslot sequence, which represent all of the possible coding choices of the
participants. In each timeslot the participants record a tuple $\left(
t,c,b\right)  $ where $t$ is the index for the timeslot (which we can simply
think of as an integer so that $t=1,2,3,\ldots,N$), $c$ is the bit
representing the choice of coding basis (we adopt the convention that
$X\equiv0$ and $Y\equiv1$) and $b$ is the actual received/transmitted bit
value. Thus in a timeslot $t$ we have the participants recording the following tuples:%

\begin{align*}
\text{Alice }  &  \text{: }\left(  t,c_{A},b_{A}\right) \\
\text{Relay 1 }  &  \text{: }\left(  t,c_{1},b_{1}\right) \\
\text{Relay 2 }  &  \text{: }\left(  t,c_{2},b_{2}\right) \\
\text{Bob }  &  \text{: }\left(  t,c_{B},b_{B}\right)
\end{align*}
The partitioned channels we have described above are generated by the possible
values of the bit string $c_{A}c_{1}c_{2}c_{B}$. The 8 partitioned channels in
the above table are given by the possible values of the bit string
$0c_{1}c_{2}c_{B}$. In ideal operation we therefore have:%

\begin{align*}
c_{A}  &  =c_{1}~~\Rightarrow~~b_{A}=b_{1}\\
c_{1}  &  =c_{2}~~\Rightarrow~~b_{1}=b_{2}\\
c_{2}  &  =c_{B}~~\Rightarrow~~b_{2}=b_{B}%
\end{align*}
which simply states that if adjacent participants in the channel choose the
same coding basis they will record the same bit value. In other words, if
adjacent users choose the same coding basis then they could establish a
quantum key on that link, if they desired. We describe these links as `open'
and links in which adjacent users have chosen a different coding basis as
`closed'. Using the symbol $\square$ to denote an open link and the symbol
$\blacksquare$ to denote a closed link the 8 channels in the above table can
be rewritten as%

\begin{align*}
1  &  :X\square X\square X\square X\\
2  &  :X\square X\square X\blacksquare Y\\
3  &  :X\square X\blacksquare Y\blacksquare X\\
4  &  :X\square X\blacksquare Y\square Y\\
5  &  :X\blacksquare Y\blacksquare X\square X\\
6  &  :X\blacksquare Y\blacksquare X\blacksquare Y\\
7  &  :X\blacksquare Y\square Y\blacksquare X\\
8  &  :X\blacksquare Y\square Y\square Y
\end{align*}
Consider channels 2 and 5. We can see that a quantum key can only propagate
between $A$, $R_{1}$ and $R_{2}$ in timeslots described by channel 2, and only
between $R_{2}$ and $B$ in timeslots described by channel 5. However, Alice
and Bob are only interested in establishing a shared bit at the end of the
process and so $R_{2}$ can `repair' the broken link by correlating a timeslot
from channel 2 configurations with another timeslot from channel 5
configurations such that the same bit value is propagated. For example,
suppose in timeslots 5 and 12 we have these configurations, respectively, then
we might have the recorded tuples%
\[%
\begin{tabular}
[c]{c|cccc}
& Alice & $R_{1}$ & $R_{2}$ & Bob\\\hline
Channel 2 & $\left(  5,0,1\right)  $ & $\left(  5,0,1\right)  $ & $\left(
5,0,1\right)  $ & $\left(  5,1,0\right)  $\\
Channel 5 & $\left(  12,0,1\right)  $ & $\left(  12,1,0\right)  $ & $\left(
12,0,1\right)  $ & $\left(  12,0,1\right)  $%
\end{tabular}
\]
The relay $R_{2}$\ simply needs to announce to Alice and Bob that Alice should
use the bit value from timeslot 5 and Bob should use the bit value from
timeslot 12. Thus 2 partially closed channels are effectively combined into
one open channel and Alice and Bob can establish a shared bit value by
combining their data in timeslots 5 and 12. Alice ignores her recorded data in
timeslot 12 and Bob ignores his recorded data in timeslot 5.

For each partially closed channel in the above there is a \textit{dual}
channel which can be used to correlate the data so that a shared bit value can
be established with the assistance of the intermediate relays. There are 6
partially closed channels, one fully open channel ($X\square X\square X\square
X$) and one fully closed channel ($X\blacksquare Y\blacksquare X\blacksquare
Y$). Thus by transporting bits across broken links in this fashion we can see
that Alice and Bob can establish $N/2$ shared bits, on average, just as they
would for single link QKD.

If a partitioned channel is described by the bit string $c_{A}c_{1}c_{2}c_{B}$
then the bit string for the dual channel is given by%
\[
c_{A}c_{1}c_{2}c_{B}\oplus c_{A}\bar{c}_{A}c_{A}\bar{c}_{A}%
\]
where $\oplus$ is the bitwise exclusive-or of the bit strings and $\bar{c}$ is
the bit complement of $c$. This can be extended to the situation with $n$ relays:%

\[
A\longrightarrow\fbox{$R_1$}\longrightarrow\ldots\longrightarrow\fbox{$%
R_n$}\longrightarrow B
\]
where we can partition the data into $2^{n}$ channels distributed randomly
across the sequence of timeslots. If one of these partitions is described by
the bit string $c_{A}c_{1}c_{2}c_{3}\ldots c_{n-2}c_{n-1}c_{n}c_{B}$ then its
dual partition is described by%
\begin{align*}
&  c_{A}c_{1}c_{2}c_{3}\ldots c_{n-2}c_{n-1}c_{n}c_{B}\oplus c_{A}\bar{c}%
_{A}c_{A}\bar{c}_{A}\ldots c_{A}\bar{c}_{A}c_{A}\bar{c}_{A}\text{ \ \ }\left(
n\text{ even}\right) \\
&  c_{A}c_{1}c_{2}c_{3}\ldots c_{n-2}c_{n-1}c_{n}c_{B}\oplus c_{A}\bar{c}%
_{A}c_{A}\bar{c}_{A}\ldots\bar{c}_{A}c_{A}\bar{c}_{A}c_{A}\text{ \ \ }\left(
n\text{ odd}\right)
\end{align*}
It is clear that to each partially closed channel there is a single unique
dual channel (also partially closed). The fully-open and fully closed
partitions are, of course, duals of one another. Hence we can see that if
there are $N$ timeslots, then, even with $n$ relays, we can use this bit
transport technique over dual partitions to establish a shared key between
Alice and Bob that is, on average, $N/2$ bits in length.

Bit transport, as outlined above, is a classical \textit{post-processing}
technique (although it \textit{can} be performed during the quantum key
transmission). When seen in this perspective we note that we can think of the
entire quantum transmission as a series of $N$ sequential experiments. It is
the post-processing of the data that allows us to establish a shared secret
key from this data. This is true of single link QKD as it is for relay-based
schemes. This change of perspective allows us to consider new network
operations such as asynchronous quantum key exchange [8]. Indeed, by
considering the data collected on the quantum transmissions simply as `data'
we might also wish to use the network relay nodes as a sensitive network
monitoring tool where problems with particular paths or links can be detected
and communications re-routed. There is no reason why we have to use the
recorded quantum data only in security applications.

\subsection{QKD with Bit Revelation}

As an illustration of the application of this perspective of a QKD channel as
a series of sequential experiments coupled with a post-processing technique,
we consider the possibility of key establishment over a quantum channel in
which the bit values are publicly revealed. In the standard operation of QKD
the basis information is revealed and the bit values are kept secret. We show
that it is also possible to establish a shared secret key by revelation of the
bit values, but keeping the basis information secret.

Assuming a lossless channel and ideal detections Alice and Bob will each
possess a set of $N$ tuples $\left\{  \left(  t,c_{A},b_{A}\right)  \right\}
$ and $\left\{  \left(  t,c_{B},b_{B}\right)  \right\}  $, respectively. Each
of Alice's tuples contains 2 secret bits of information; the basis bit $c_{A}$
and the coded bit value $b_{A}$. In the standard BB84 protocol\ [1] the basis
information is publicly revealed and a sifting process employed to select only
those tuples, for the same $t$, where $c_{A}=c_{B}$. Public examination of a
small random sample of these sifted tuples reveals (under our assumption of
ideal conditions) the presence of an eavesdropper. The possible transmissions
and outcomes on a QKD channel operating the BB84 protocol are shown in the
table below in terms of the recorded tuples $\left(  t,c,b\right)  $ for a
given timeslot labelled by $t\bigskip$%

\begin{tabular}
[c]{ccc}%
\underline{Alice} & \underline{Bob} & \underline{Probability}\\
&  & \\
$\left(  t,0,0\right)  _{A}$ & $\left(  t,0,0\right)  _{B}$ & 1\\
& $\left(  t,1,0\right)  _{B}$ & 1/2\\
& $\left(  t,1,1\right)  _{B}$ & 1/2\\
&  & \\
$\left(  t,0,1\right)  _{A}$ & $\left(  t,0,1\right)  _{B}$ & 1\\
& $\left(  t,1,0\right)  _{B}$ & 1/2\\
& $\left(  t,1,1\right)  _{B}$ & 1/2\\
&  & \\
$\left(  t,1,0\right)  _{A}$ & $\left(  t,1,0\right)  _{B}$ & 1\\
& $\left(  t,0,0\right)  _{B}$ & 1/2\\
& $\left(  t,0,1\right)  _{B}$ & 1/2\\
&  & \\
$\left(  t,1,1\right)  _{A}$ & $\left(  t,1,1\right)  _{B}$ & 1\\
& $\left(  t,0,0\right)  _{B}$ & 1/2\\
& $\left(  t,0,1\right)  _{B}$ & 1/2
\end{tabular}
\bigskip\newline where the probability refers to the probability that, for the
transmitted state, Bob records that tuple given his measurement basis. At this
point everything is per the usual BB84 protocol and neither basis nor bit
information has been revealed publicly. Let us consider the case where Alice
chooses to reveal, for a given timeslot, the bit value $b_{A}$, whilst keeping
the basis bit $c_{A}$ secret. We shall suppose that in this timeslot she has
recorded the tuple $\left(  t,0,0\right)  _{A}$ so that her revealed bit value
is $b_{A}=0$. If, in this timeslot, Bob has chosen a basis value of $c_{B}=0$,
then he is using the same basis as Alice and he will have recorded the triple
$\left(  t,0,0\right)  _{B}$ with unit probability. With Alice's announcement
he knows that he has measured the correct bit value, but he cannot tell
whether Alice's tuple is $\left(  t,0,0\right)  _{A}$ or $\left(
t,1,0\right)  _{A}$. He knows that with his chosen basis and the measurement
result his tuple is more likey (with a probability of 2/3) to have been a
result of Alice's tuple $\left(  t,0,0\right)  _{A}$, but he cannot be
certain. In other words, he cannot unambiguously decode his measurement to
yield Alice's secret basis value if his recorded tuple is $\left(
t,0,0\right)  _{B}$.

If he has chosen a basis value of $c_{B}=1$ then his possible tuples resulting
from this choice and Alice's input tuple are $\left(  t,1,0\right)  _{B}$ and
$\left(  t,1,1\right)  _{B}$. If he records the tuple $\left(  t,1,0\right)
_{B}$ then his bit value is in agreement with Alice's revealed bit but he
cannot tell (with equal probability) whether this has arisen from an input
tuple of $\left(  t,0,0\right)  _{A}$ or $\left(  t,1,0\right)  _{A}$.
However, if his measurement result yields the tuple $\left(  t,1,1\right)
_{B}$ then he knows with certainty that, with Alice's revealed bit value of
$b_{A}=0$ this could only have arisen from an input tuple of $\left(
t,0,0\right)  _{A}$ and so he knows that Alice's basis bit $\left(
c_{A}=0\right)  $ is the complement of his basis bit $\left(  c_{B}=1\right)
$. Similar considerations apply for each of Alice's possible input tuples so
that we can establish the following adapted protocol for quantum key
establishment using the BB84 protocol with bit revelation :

\begin{enumerate}
\item Alice chooses a coding basis at random

\item Alice chooses a state from that basis at random

\item For each timeslot Alice transmits her chosen quantum state to Bob

\item Alice records the tuple $\left(  t,c_{A},b_{A}\right)  $ where $t$ is
the timeslot, $c_{A}$ is the coding basis and $b_{A}$ is the bit value

\item For each timeslot Bob chooses, at random, a coding basis in which to
decode (i.e. measure) the transmitted quantum state

\item Bob records the tuple $\left(  t,c_{B},b_{B}\right)  $ where $c_{B}$ is
Bob's decoding (measurement) basis and $b_{B}$ is the decoded (measured) bit

\item For each timeslot Alice reveals her value $b$ and Bob compares this with
his value $b_{B}$ and they discard those timeslots in which $b_{A}=b_{B}$

\item Bob determines the logical complement $\overline{c_{B}}$

\item Alice and Bob now have a list in which $t,c_{A}$ and $\overline{c_{B}}$
are in agreement. A random sample of these results are chosen and the values
$c$ and $\overline{c_{B}}$ compared. This gives an estimate of the error rate
for the channel. These compared timeslots are discarded. The remaining values
$c_{A}=\overline{c_{B}}$ can be used as a key

\item The remaining timeslots can be renumbered for convenience so that
$t\in\left\{  1,2,\ldots,n\right\}  $ where $n$ is the total number of
successful timeslots
\end{enumerate}

This protocol variant of BB84 discards (on average) 5/8 of the transmitted
timeslots and is worse than the standard protocol in this regard. However, it
offers a slight advantage over the standard BB84 protocol in that an
eavesdropper employing a basic intercept/resend strategy causes a higher error
rate on the key data.

Whilst we would not suggest adopting this protocol over the standard operation
of BB84 we note that if the eavesdropper employs a strategy that is optimised
for the BB84 protocol with basis revelation then it may not be optimsed for
this adapted BB84 protocol with bit revelation. As both of these protocols
rely on \textit{post-processing }techniques Alice and Bob can select, at
random, which protocol to operate for a given recorded timeslot. Thus an
eavesdropper needs to optimise any strategy over both protocol variants.
Indeed, in general, an eavesdropper needs to optimise over all possible
post-processing protocols that Alice and Bob can employ.

Here, of course, we have not undertaken a full security analysis of this
alternative protocol. If such a protocol were ever required to be used in
practice then a full security analysis is necessary, but it would unduly
distract us from the main point we are making. Here we are using this as an
illustration of the flexibility of considering a QKD channel as a quantum
`experiment' which simply collects data. It is the \textit{post-processing} of
this data that gives us the possibility of bit transport and of randomizing
over different choices of post-processing protocol.

\section{QKD Without Public Bit Comparison}

In order to detect errors without any public bit comparison Alice and Bob must
operate a \textit{duplex} QKD channel. That is, Alice transmits photons to
Bob, according to the BB84 protocol, and Bob transmits photons to Alice
according to the BB84 protocol. For convenience we shall imagine these to be
interleaved so that for odd timeslots a photon is transmitted by Alice whereas
Bob transmits in even timeslots. These can, in fact, be two \textit{entirely
separate} transmissions; all that is required is that we can uniquely
correlate a particular transmission with a particular measurement, which we
can achieve using the technique of bit transport. We can view the duplex
channel as being of the form $A\rightarrow R\rightarrow B$ folded back on
itself. As above, a full security analysis of this protocol would unduly
detract from the point we are making here; the duplex channel provides us with
an example of another capability unlocked by applying post-processing
techniques such as bit-transport.

It is best to illustrate the technique with an example. The table below gives
an example duplex transmission over 18 timeslots. We assume that each timeslot
is occupied and that each photon reaches its destination. This is, of course,
not true in practice, but it is easy to accommodate timelots where nothing is
transmitted or received. We use a tilde to denote values in even timeslots
(transmissions from Bob to Alice) so that $\tilde{b}_{B}$, for example, would
represent Bob's transmitted bit value in one of these even timeslots. We also
use the symbol $\blacklozenge$ to denote the bit in situations where Alice and
Bob choose a different coding basis. For clarity we have also reverted to the
description of the coding basis using $X$ or $Y$ rather than using the bit
value to denote this choice.%

\[%
\begin{tabular}
[c]{|c|c|c|c|c|c|c|c|c|c|c|c|c|c|c|c|c|c|c|}\hline
Timeslot & 1 & 2 & 3 & 4 & 5 & 6 & 7 & 8 & 9 & 10 & 11 & 12 & 13 & 14 & 15 &
16 & 17 & 18\\\hline
$c_{A}$ & $X$ &  & $X$ &  & $Y$ &  & $X$ &  & $Y$ &  & $X$ &  & $Y$ &  & $Y$ &
& $X$ & \\\hline
$b_{A}$ & 1 &  & 1 &  & 0 &  & 0 &  & 1 &  & 1 &  & 0 &  & 1 &  & 0 & \\\hline
$c_{B}$ & $Y$ &  & $X$ &  & $Y$ &  & $Y$ &  & $Y$ &  & $X$ &  & $X$ &  & $Y$ &
& $Y$ & \\\hline
$b_{B}$ & $\blacklozenge$ &  & 1 &  & 0 &  & $\blacklozenge$ &  & 1 &  & 1 &
& $\blacklozenge$ &  & 1 &  & $\blacklozenge$ & \\\hline
$\tilde{c}_{B}$ &  & $X$ &  & $X$ &  & $Y$ &  & $X$ &  & $X$ &  & $Y$ &  & $Y$
&  & $Y$ &  & $X$\\\hline
$\tilde{b}_{B}$ &  & 0 &  & 0 &  & 1 &  & 1 &  & 1 &  & 0 &  & 0 &  & 1 &  &
0\\\hline
$\tilde{c}_{A}$ &  & $X$ &  & $Y$ &  & $Y$ &  & $X$ &  & $Y$ &  & $X$ &  & $Y$
&  & $Y$ &  & $X$\\\hline
$\tilde{b}_{A}$ &  & 0 &  & $\blacklozenge$ &  & 1 &  & 1 &  & $\blacklozenge$
&  & $\blacklozenge$ &  & 0 &  & 1 &  & 0\\\hline
\end{tabular}
\]
Alice informs Bob of her basis choices for both transmission and measurement.
Bob filters this data into 3 sets. The first is the data for which they expect
no agreement because they have chosen different bases. In the table this first
set consists of timeslots $t=1,4,7,10,12,13,17$. Bob informs Alice of these
timeslots and they are discarded. The second and third sets consist of the
remaining odd and even timeslots, respectively. In a perfect world and in the
absence of an eavesdropper, Alice and Bob should have the same recorded tuples
for sets 2 and 3.

In order to check this agreement Bob chooses a timeslot from set 2 and reads
the measured bit value $b_{B}$ from the recorded tuple. He then looks for an
element of set 3 in which his transmitted bit value $\tilde{b}_{B}=b_{B}$. He
sends Alice the $t$ value for these timeslots. Alice compares her bit values
$\tilde{b}_{A}$ and $b_{A}$ from these two timeslots. They should be equal. If
there are errors on the channel, caused by an eavesdropper or practical
imperfections, then there is a finite probability that Alice's comparison will
fail. If Bob transmits a sufficient number of these timeslot pairs from sets 2
and 3 then the probability of an error remaining undetected can be made
negligibly small.

Of course, by revelation of which timeslots have equal values, Bob has leaked
information to any eavesdropper. If the eavesdropper has measured both
channels using a standard intercept/resend strategy in the coding bases, then
if only one of her basis guesses were correct she would know the bit value for
both channels $A\rightarrow B$ and $B\rightarrow A$. Instead of searching for
an identical bit value from sets 2 and 3, Bob could simply look at sequential
tuples and transmit an extra bit of information that tells Alice whether or
not to perform a bit flip on her recorded bit values from set 3. Using the
example data from the table we can see that these sets lead to the following
tuples recorded by Bob%

\[%
\begin{tabular}
[c]{|c|c|cc|c|c|}\cline{1-2}\cline{5-6}%
\multicolumn{2}{|c|}{Set 2} &  &  & \multicolumn{2}{|c|}{Set 3}\\\cline{1-2}%
\cline{5-6}%
Timeslot & $b_{B}$ &  &  & Timeslot & $\tilde{b}_{B}$\\\cline{1-2}\cline{5-6}%
3 & 1 &  &  & 2 & 0\\\cline{1-2}\cline{5-6}%
5 & 0 &  &  & 6 & 1\\\cline{1-2}\cline{5-6}%
9 & 1 &  &  & 8 & 1\\\cline{1-2}\cline{5-6}%
11 & 1 &  &  & 14 & 0\\\cline{1-2}\cline{5-6}%
15 & 1 &  &  & 16 & 1\\\cline{1-2}\cline{5-6}%
$-$ & $-$ &  &  & 18 & 0\\\cline{1-2}\cline{5-6}%
\end{tabular}
\]
After the quantum transmission (or indeed during the quantum transmission, if
so desired) Bob sends Alice a list of tuples $\left(  t,\tilde{t},f\right)  $
where $f\in\left\{  0,1\right\}  $ such that 0 means `no flip' and 1 means
`flip'. For example, if Bob sends Alice the tuple $\left(  3,2,1\right)  $
this means that Alice is to take her data from timeslots 2 and 3 and flip the
recorded bit she has obtained in timeslot 2. We could, equally, require the
data from Alice's transmission to be flipped instead of the data from Bob's
transmission; all that is needed is that one of the 2 bits from set 2 or set 3
are flipped (or not flipped if they should be in agreement). The extra bit
that Bob sends is therefore a parity check bit. Thus in the example
transmission above Bob would send Alice the following list of tuples%
\begin{align*}
&  \left(  3,2,1\right) \\
&  \left(  5,6,1\right) \\
&  \left(  9,8,0\right) \\
&  \left(  11,14,1\right) \\
&  \left(  15,16,0\right)
\end{align*}
The parity bit revealed by Bob is, of course, an extra source of potentially
useful information to an eavesdropper. To eliminate this information gain by a
passive Eve, Alice and Bob could adopt the rule that the compared timeslots
must be understood as follows. If the recorded bit values $\left(  b,\tilde
{b}\right)  $ are $\left(  0,1\right)  $ or $\left(  0,0\right)  $ then this
is taken to be a 0 for the final key. If the recorded bit values $\left(
b,\tilde{b}\right)  $ are $\left(  1,0\right)  $ or $\left(  1,1\right)  $
then this is taken to be a 1 for the final key. An \textit{active}
eavesdropper performing, for example, a standard intercept/resend strategy in
the coding bases for every timeslot does gain extra information, of course.
However, Eve's intervention in this case causes an error rate of 3/8.

The actual bit values recorded by Alice and Bob are never publicly revealed,
although 1 bit of information, the parity bit $f$, is revealed. The
fundamental difference between this protocol and BB84 operated in single link
QKD mode, is that the bit transport mechanism allows Alice and Bob to use
their \textit{entire filtered transmission} to check for errors, rather than
just a random sample which is then discarded. Alice and Bob, therefore have
access to their entire data set, without compromising their final key, to
gather information about what is happening on the channel between them. Of
course they could, if they choose, also perform a standard random sampling on
the data from sets 2 and 3 (discarding timeslots in which bit values are
publicly revealed).

In practice, the duplex technique would not be used to establish a key since
single link QKD can already be operated in such a fashion as to give
unconditional security [11-14]. However, one could imagine situations where
this duplex technique may be useful to obtain extra information about the
channel. Indeed, given that we would envision that in any practical network
both Alice and Bob will have both detection and transmission capability then
it costs little extra to perform duplex transmissions anyway. In the next
section we consider a relay using a different technique employing GHZ-type
states which performs a kind of automatic bit transport.

\section{Relay Key Distribution with GHZ-type States}

In the preceding discussion we have assumed a key distribution scheme
employing a BB84 protocol and the transmission of spin-1/2 states. This
combination is one of the most robust practically and can easily be achieved
with quasi-single photon sources and optical fibres. Indeed, commercial QKD
systems employing these mechanisms have been available for a number of years.
It could be argued that the current explosion of interest in quantum
information processing was kick-started by the development of QKD. It is clear
that whilst QKD based on quasi-single photon sources is a robust and practical
technology, it is only the tip of the iceberg as far as the possibilities for
the exploitation of quantum mechanics in information processing. With
increasing advances in the control and manipulation of entangled sources we
expect to see many new and fascinating technologies emerge in the not too
distant future. With this in mind, therefore, we consider one application of
intermediate nodes on a channel that is currently practically infeasible.

In this application the intermediate node is not really acting as a relay, as
such, but it is interesting because the GHZ-type correlation automatically
achieves some of the features of bit transport. Let us consider an
intermediate node between Alice and Bob who is able to prepare spin-1/2 states
in the correlated form (and we use a binary notation for the state here,
rather than our previous $\left\vert \pm\right\rangle $)%

\[
\left\vert \psi\right\rangle =\frac{1}{\sqrt{2}}\left(  \left\vert
B_{1},0\right\rangle +\left\vert B_{3},1\right\rangle \right)
\]
where $\left\vert B_{1}\right\rangle $ and $\left\vert B_{3}\right\rangle $
are Bell states given by%

\begin{align*}
\left\vert B_{1}\right\rangle  &  =\frac{1}{\sqrt{2}}\left(  \left\vert
00\right\rangle +\left\vert 11\right\rangle \right) \\
& \\
\left\vert B_{3}\right\rangle  &  =\frac{1}{\sqrt{2}}\left(  \left\vert
01\right\rangle +\left\vert 10\right\rangle \right)
\end{align*}
One of the particles in the Bell states is sent to Alice and its partner to
Bob whilst the intermediary holds on to the remaining particle. A number of
such particles are sent to Alice and Bob in well-defined timeslots. All 3
particles are stored (technology permitting, of course) for future use. When
Alice and Bob wish to exchange a secret key they inform the intermediary who
then makes a measurement of the spin variable of his stored particle. The
intermediary publishes this list, which is just a random binary bit string, to
Alice and Bob over a public channel.

Alice and Bob, for each particle, then make a spin measurement. The
measurement of the intermediary will have projected their particles into
either a correlated state or an anti-correlated state. Accordingly, with the
random binary bit string sent by the intermediary, Alice and Bob can now
establish a shared secret random bit string from their measurements. For
example, if Alice measures a bit value 1 and this is associated with a bit
value of 1 from the bit string sent by the intermediary, then she knows that
Bob will have measured the bit value 0. They can establish a key from the
inferred parity of their measurements. The interesting feature of this scheme
is that Alice and Bob need not communicate directly, nor need they perform
their measurements synchronously (assuming secure storage). Furthermore, they
can check whether their particles are indeed correlated (after the measurement
of the intermediary) by choosing a random sample and checking to see whether
the Bell inequality is violated, although this does require some public
communication. Another useful feature of this scheme is that all of the
transmitted particles are used to establish the key.

As it stands, this scheme is vulnerable to a meet-in-the-middle attack by an
eavesdropper who need only intercept both particles sent to Alice and Bob,
perform a Bell measurement to determine whether the state is correlated or
anti-correlated, and re-transmit the particles in a corresponding Bell state
to Alice and Bob. The Bell measurement of the eavesdropper will project the
intermediary's spin into the state $\left\vert 0\right\rangle $ or $\left\vert
1\right\rangle $ at random. In order to frustrate this attack, the
intermediary randomizes the transmitted particles over the timeslots. Thus, if
the intermediary prepares 5 states for transmission to Alice and Bob, then
Alice's particle from state 1 could be sent in timeslot 1, with the particle
sent to Bob in timeslot 1 coming from, say, state 4. When Alice and Bob wish
to establish a key the intermediary publishes the random binary string
representing the spin measurements made, and also a list of tuples containing
the information about which timeslots must be associated together. Again we
see the key idea underlying bit transport, that of an intermediary correlating
disparate data sets, finding an application here.

In a functional sense, this scheme is a kind of quantum version of the
Needham-Schroeder protocol [15] for the establishment of keys. The
Needham-Schroeder protocol is at the heart of the Kerberos protocol for the
exchange of symmetric keys. Kerberos, and other classical key management
protocols, will assume a central importance in the advent of a working quantum
computer able to tackle problems of significant input size. We can see that
this quantum protocol is much neater in principle than the classical protocols
which require several challenge-response communications between the parties.
Of course such quantum protocols are beyond the reach of existing technology,
but the technology advances required to construct a quantum computer of
significant capability may also be instrumental in allowing us to practically
implement the more speculative protocols involving GHZ-type states such as the
one we have outlined here.

In this application, the intermediary must be trusted. This is also true of
the IR relays described above when acting to extend the distance for QKD.
Indeed, in the QKD distance extension application, as described above, if Eve
manages to compromise any \textit{one} relay, then she will compromise the
entire communication. An ingenious scheme [9] using distinct physical paths on
a network, together with secret sharing, can alleviate some of this trust
burden. As we now discuss, this technique can also be adapted to work on
single dedicated paths containing relays.

\section{Trusting the Relays}

One of the most attractive features of QKD is its promise of unconditional
security, when correctly implemented to overcome any side-channel attacks
(see, for example, [16]). If we need to add extra \textit{trusted} relays in
order to operate the technique over the required network distances, then from
a security perspective, the benefit of QKD over conventional key distribution
systems becomes much harder to argue. The protocol operation we have so far
discussed means that an attacker only has to compromise \textit{one} relay in
a channel in order to access the final key. Using multi-path networks,
together with secret-sharing, can alleviate this problem somewhat [9]. In
essence, the distinct network paths between Alice and Bob can be thought of as
shares and the final key obtained by the binary addition of the keys obtained
on the different paths. An attacker, therefore, has to compromise all paths in
order to obtain the final key.

We can achieve the same functionality over a single path containing relays by
creating distinct \textit{logical} channels on that path. This is best
illustrated by an example. We suppose that our two end users, Alice and Bob,
require 3 relays to effectively span the distance between them and establish a
quantum key. The path is, therefore, of the form%

\[
A\longrightarrow\fbox{$R_1$}\longrightarrow\fbox{$R_2$}\longrightarrow
\fbox{$R_3$}\longrightarrow B
\]
Let us add an \textit{additional} relay spaced such that any 3 of the 4 relays
are sufficient to span the distance for the purposes of quantum key
establishment. With a suitable re-labelling of the relays the channel is now
of the form%

\[
A\longrightarrow\fbox{$R_1$}\longrightarrow\fbox{$R_2$}\longrightarrow
\fbox{$R_3$}\longrightarrow\fbox{$R_4$}\longrightarrow B
\]
Now let us suppose that we operate the relays such that for any transmission
timeslot they `drop out' at random. By drop out here we mean that they allow
the quantum signal to pass through unmeasured and unaffected. If 2 or more
drop out in any one timeslot a quantum key cannot be established on the path
because the quantum signal will not span the distance. If the probability that
any one relay drops out is $p$, and we assume independent relays, then the
probability that any timeslot will result in a successful transmission is%

\begin{equation}
P\left(  \text{open}\right)  =\left(  1-p\right)  ^{4}+4p\left(  1-p\right)
^{3}%
\end{equation}
where we describe such a path as `open'. For $p=1/2$ we see that 5/16 of the
timeslots, on average, will lead to an open path. However, what Alice and Bob
require is to use only timeslots in which precisely 3 out of the 4 relays are
operational. This is, as we shall see, because they are going to establish
their final key by binary addition of the keys established over the separate
logical channels in which precisely 3 out of the 4 relays are operational.
Thus the fraction of useful timeslots on average, for Alice and Bob, is given by%

\begin{equation}
f=4p\left(  1-p\right)  ^{3}%
\end{equation}
which for $p=1/2$ is 1/4. The useful timeslots for Alice and Bob will be those
in which the transmission has been effected by one of the following relay configurations%

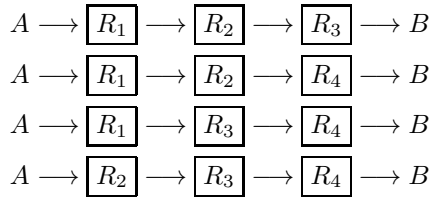
\begin{align*}
A  &  \longrightarrow\fbox{$R_1$}\longrightarrow\fbox{$R_2$}\longrightarrow
\fbox{$R_3$}\longrightarrow B\\
A  &  \longrightarrow\fbox{$R_1$}\longrightarrow\fbox{$R_2$}\longrightarrow
\fbox{$R_4$}\longrightarrow B\\
A  &  \longrightarrow\fbox{$R_1$}\longrightarrow\fbox{$R_3$}\longrightarrow
\fbox{$R_4$}\longrightarrow B\\
A  &  \longrightarrow\fbox{$R_2$}\longrightarrow\fbox{$R_3$}\longrightarrow
\fbox{$R_4$}\longrightarrow B
\end{align*}
If we denote the keys established for each of these logical channels as
$QK_{ijk}$ then Alice and Bob will establish their final quantum key $QK_{AB}$
by the simple expedient of performing the binary addition for these separate
quantum keys so that%

\begin{equation}
QK_{AB}=QK_{123}\oplus QK_{124}\oplus QK_{134}\oplus QK_{234}%
\end{equation}
Each key is a share of the final key and only those participants with access
to all shares, in this case Alice and Bob, can recover the final secret key.
An eavesdropper with access to the information of only one relay cannot
establish any information about the final key. An eavesdropper with full
control of one relay might decide to disable the drop out feature so that it
is always operational. However, this can easily be detected by the legitimate
network participants by examination of the quantum data since the relay is
then effectively acting as an eavesdropper performing a standard
intercept/resend strategy in each timeslot. In order to compromise the entire
channel the eavesdropper needs to compromise \textit{every} relay on the
channel. In other words, in order to establish a quantum key, Alice and Bob
need only trust \textit{one} relay on the channel.

If we couple this technique with the use of distinct physical paths on the
network, then we can see that in order to compromise the key $QK_{AB}$ an
eavesdropper needs to compromise \textit{all} of the relays on the network
that could be involved in the transmission between Alice and Bob. Whilst this
is theoretically possible, it is a considerable practical challenge for any
eavesdropper. If we also employ bit transport to enable \textit{asynchronous}
quantum key establishment [8] then not only must an eavesdropper compromise
all the relays on the channel but must compromise those relays and collect
data on all possible paths between Alice and Bob since the inception of the
network. A considerable practical challenge indeed.

The technique outlined above can clearly be extended to any number of relays.
The important feature is that if we have just enough relays to span the
distance then the entire channel is vulnerable to the compromise of a single
relay. If we add just one extra relay, superfluous to our distance spanning
requirement, then we can operate the relays in such a way as to ensure that
the channel can only be compromised if every relay on the channel is
compromised. There is a price, however. If we need $n-1$ relays to span the
distance then with the addition of an extra relay and the drop out operation
described above, we would only have a fraction of useful channels for Alice
and Bob given by%

\begin{equation}
f=np\left(  1-p\right)  ^{n-1}%
\end{equation}
Thus if we require 9 relays to span the distance, the addition of a single
relay operated as above with $p=1/2$ means that $f\approx0.01$ so that, on
average, approximately only 1 out every 100 timeslots is useful to Alice and
Bob in forming the final key, and this is on the assumption that every
timeslot contains transmissions that reach their destination.

The operation of the relay channel described above is not, of course, the only
way to operate such channels, nor are we limited to adding in only one extra
relay. All that is required for Alice and Bob to be able to establish a final
key is that

\begin{description}
\item[(i)] Alice and Bob participate in all logical channels

\item[(ii)] They ensure that every relay is absent from at least one of the
channels used to form the key
\end{description}

The relays thus have no knowledge of the final key established between Alice
and Bob. Indeed, from a security perspective we might wish to add a relay on a
channel, even when it is not strictly necessary, in order to access this
feature. We might, for example, know that the relays are very secure (consider
a relay placed on a satellite, for example, then practical limitations might
make us more confident that it cannot be compromised). Alice and Bob, for very
good reasons, may still not wish another party to have access to their key.
This feature is particularly important in cases where the provision of a QKD
network might be offered as a managed service by a network operator.

\section{Conclusion}

QKD is a mature and established practical technology [17] that has been
implemented in several large-scale trials (see, for example[18]) as well as
commercially. The current principal limitation on any network implementation
is the effective distance over which the technique is feasible. Overcoming
this limitation requires the use of intermediate relays which are
conventionally operated in a link-by-link mode. This reduces the
attractiveness of a network solution that we might wish to claim as providing
unconditional security.

Here, we have shown how, with a suitable adaptation, IR relays can be operated
on a QKD network to extend the distance over which successful quantum key
exchange can be performed. By introducing the technique of bit transport we
have shown that the size of final key established between Alice and Bob is
unaffected by the introduction of such relays. Bit transport is a powerful
technique for extending the capability and flexibility of QKD networks and
moves the perspective from a sychronous QKD network to that of a network-based
collection of quantum experiments in which the collected data can be used to
establish a key both synchronously or asynchronously. Furthermore, the
potential establishment of quantum keys between any two network nodes also
gives us a pool of data for monitoring purposes, quite independent of any
security application.

The possibility of duplex channels for establishing keys without public
revelation is another application of bit transport. This technique also gives
us the capability to use the entire filtered transmission to monitor the
channel without compromising the security of the final key. Some of the
features of bit transport can automatically be achieved by using more complex
entangled quantum states such as the GHZ-type state we have considered. Whilst
such schemes are currently impractical they suggest that our exploitation of
the capabilities of quantum networks is still in its infancy, although
significant progress is being made in this direction [19]. Indeed, it is fair
to say that a radical revision of our understanding of information processing
has been engendered by the exploitation of quantum mechanics (see, for
example, Shor's seminal paper on quantum computing [20] which has
revolutionised our perspective on computation. Other examples of the
implications of this shift to quantum information can be found in [21]). It
seems to us likely that further significant advances are still to be made.

Any network node, whether used for routing or to increase the distance over
which a signal can be transmitted, must be trusted. We have discussed here how
the creation of distinct logical channels using a random drop-out technique
can radically alter the trust requirements on the intermediate nodes. This
technique, when coupled with others such as multi-path QKD secret sharing and
asynchronous quantum key establishment, gives us a very powerful methodology
for the operation of a quantum key exchange network. Whilst the commercial
arguments for single link QKD may not be compelling, it is clear that
network-based QKD with the techniques outlined above becomes a commercially
more attractive proposition.\bigskip

\textbf{References}

\begin{enumerate}
\item Bennett, C.H.; Brassard, G. Quantum Cryptography: Public Key
Distribution and Coin Tossing. \textit{Proceedings of the IEEE International
Conference on Computers, Systems, and Signal Processing}, Bangalore, India,
p.175, $\left(  1984\right)  $

\item Ekert, A.K., Quantum cryptography based on Bell's theorem, \textit{Phys.
Rev. Lett}. \textbf{67}, 661--663 $(1991)$

\item Gisin, N.; Ribordy, G.; Tittel, W., Zbinden, H. Quantum Cryptography,
\textit{Rev. Mod. Phys}., \textbf{74}, 145--195, $\left(  2002\right)  $;
Scarani, V., Bechmann-Pasquinucci, H., Cerf, N.J., Du\v{s}ek, M.,
L\"{u}tkenhaus, N., Peev, M., \textit{Rev. Mod. Phys}. \textbf{81}, 1301
$(2009)$; Loepp, S. and Wootters, W. K., Protecting information: from
classical error correction to quantum key distribution, (CUP, 2006);

\item Townsend, P.D , Phoenix, S.J.D., Blow, K.J., Barnett, S.M., Design of
Quantum Cryptography Systems for Passive Optical Networks, \textit{Elect.
Lett}., \textbf{30} 1875 $(1994)$

\item Phoenix, S.J.D., Barnett, S.M., Townsend, P.D , Blow, K.J., Multi-User
Quantum Cryptography on Optical Networks" \textit{J. Mod. Opt}., \textbf{42}
1155 $(1995)$

\item Sasaki, M.; \textit{et.al.} Field test of quantum key distribution in
the Tokyo QKD Network. \textit{Optics Express}, \textbf{19 (11)}, 10387-10409,
$\left(  2011\right)  $

\item Bechmann-Pasquinucci, H.; Pasquinucci, A., Quantum Key Distribution with
Trusted Quantum Relay, $\left(  2005\right)  $\textbf{, }%
http://arxiv.org/abs/quant-ph/0505089 (accessed Mar 19, 2012)

\item Barnett, S.M., Phoenix, S.J.D , Asynchronous Quantum Key Distribution on
a Relay Network, \textit{J. Mod. Opt}., \textbf{59 (15)}, 1349-1354 $(2012)$

\item Beals, T.R.; Sanders, B.C., Distributed relay protocol for probabilistic
information theoretic security in a randomly-compromised network.
\textit{Third International Conference on Information Theoretic Security}
(ICITS), pp.29--39, $\left(  2008\right)  $

\item Barnett, S.M.; Phoenix, S.J.D., Securing a Quantum Key Distribution
Relay Network Using Secret Sharing, \textit{Proceedings of the IEEE GCC
Conference}, Dubai, UAE, pp.143-145, $\left(  2011\right)  $

\item Mayers, D., Quantum key distribution and string oblivious transfer in
noisy channels, in\textit{ Advances in Cryptology --- Proceedings of Crypto}
'96, pages 343--357, Berlin, , Springer, $\left(  1996\right)  $

\item Shor, P.W., Preskill, J., Simple proof of security of the BB84 quantum
key distribution protocol, \textit{Phys. Rev. Lett}., \textbf{85}, 441--444,
$\left(  2000\right)  $

\item L\"{u}tkenhaus, N., Security against individual attacks for realistic
quantum key distribution, \textit{Phys. Rev. A}, \textbf{61}, 052304, $(2000)$

\item Inamori, H., L\"{u}tkenhaus, N., Mayers, D., Unconditional security of
practical quantum key distribution, \textit{Eur. Phys. Journal. D},
\textbf{41}(\textbf{3}) pages 599-627, $\left(  2007\right)  $

\item See, for example, Katz, J.; Lindell, Y., \textit{Introduction to Modern
Cryptography}; Chapman \& Hall: Florida, $\left(  2008\right)  $

\item Gerhardt, I., Liu, Q., Lamas-Linares, A., Skaar, J., Kurtsiefer, C.,
Makarov, V., Full-field implementation of a perfect eavesdropper on a quantum
cryptography system, \textit{Nature Comms}, \textbf{2}, $\left(  2011\right)
$

\item Zbinden, H.; Gisin, N.; Huttner, B.; Muller, A., Tittel, W. Practical
aspects of quantum key distribution. \textit{J. Cryptology,} \textbf{11,}
1-14, $\left(  1998\right)  $

\item Sasaki, M.; \textit{et.al.} Field test of quantum key distribution in
the Tokyo QKD Network. \textit{Optics Express}, \textbf{19 (11)}, 10387-10409,
$\left(  2011\right)  $

\item Fr\"{o}hlich, B., Dynes, J.F., Lucamarini, M., Sharpe, A.W., Yuan, Z.,
Shields, A.J., A quantum access network, \textit{Nature} \textbf{501}, 69--72,
$\left(  2013\right)  $

\item Shor, P.W., Polynomial-Time Algorithms for Prime Factorization and
Discrete Logarithms on a Quantum Computer. \textit{SIAM J. Comput}.,
\textbf{26(5)}, 1484--1509, $\left(  1997\right)  $

\item Barnett, S.M., \textit{Quantum Information}; Oxford University Press:
Oxford, $\left(  2009\right)  $
\end{enumerate}

\end{document}